\pdfoutput=1

\documentclass[11pt]{article} 

\usepackage{amsmath} 
\usepackage{amsthm} 
\usepackage{amssymb}	
\usepackage{graphicx} 
\usepackage{multicol} 
\usepackage[dvips,letterpaper,margin=0.75in,bottom=0.5in]{geometry}
\makeatletter 
\renewcommand{\maketitle} 
{ \begingroup \vskip 10pt \begin{center} \large {\bf \@title}
	\vskip 10pt \large \@author \hskip 20pt \@date \end{center}
  \vskip 10pt \endgroup \setcounter{footnote}{0} }
\makeatother 
 
\newcommand{\ket}[1]{\left| #1 \right>} 
\newcommand{\braket}[2]{\left< #1 \vphantom{#2} \right|
 \left. #2 \vphantom{#1} \right>} 
\let\baraccent=\= 
\renewcommand{\=}[1]{\stackrel{#1}{=}} 

\theoremstyle{definition}

\theoremstyle{remark}

\usepackage[utf8]{inputenc} 

\usepackage{graphicx} 

\title{On the dynamics of unobserved Universe}
\author{Stéphane A. Bronoff}
\date{} 

\begin{document}
\maketitle
\abstract{This article introduces \textit{Universal Quantum Relativity} which is a simple \textit{Theory of Everything}. It relies on an ultimate doctrine that is the \emph{absence of absolute existence}. This generalizes relativity principles up to a mother quantum theory. Then it is applied to the Universe as a unique inseparable quantum system, without external observer, using a no-collapse interpretation of quantum mechanics. Assuming infinite divisibility suffices to explain through a single model the theories of all fundamental interactions: general relativity for gravitation, and the standard model of elementary particle physics for electromagnetism, the weak and strong nuclear interactions.}

\section{Introduction and Motivation}

Theoretical physics is attempting to unify quantum mechanics with general relativity in a self-consistent manner for many years. There are a number of proposed quantum gravity theories and other so-called theories of everything. Currently, there is still no complete and consistent theory, and the candidate models still need to overcome major formal and conceptual problems. Most advanced developments are String theory and Loop quantum gravity. String theory claims to unify all interactions. However it is not a background independent genuine quantum theory and it fails to give testable prediction so far. Loop quantum gravity aims just to be quantum theory of gravity. Hence it does not give a consistent picture regarding cosmology and the unification of the forces. Moreover both String theory and Loop quantum gravity does not address the remaining conceptual problems of quantum mechanics related to its interpretation. This is certainly a major drawback as \textit{locality} and certain intuitive features of \textit{realism} have been recently ruled out by experiments violating Bell's and Leggett's inequalities. 

The motivation of the present work has its root in the belief that Quantum Physics is more fundamental than in Bohr's Copenhagen interpretation. It develops the idea that the entire Universe is quantum and that no external reality,  independent of observation, exists. This suggests that the phenomena of quantum mechanics as well as of gravity, possess no properties corresponding to the idea of absolute existence. This  is simply the Einstein's relativity principle in an universal formulation.  Nothing exists by itself, that is to say in relation with anything other than itself. In the absence of something else what \lq\lq{}is\rq\rq{}, \lq\lq{}is not\rq\rq{}. This ontological dichotomy may be interpreted as a superposition of quantum states. Thus it imposes to consider the entire universe as a quantum system and to regard the quantum state as the fundamental entities. Physics was regularly pushed to re-evaluate the role of the observer banishing absolutes of the previous physics. Galilee and Einstein have shown that \textit{all that is} were observations, obtained by observers in different frames of reference, linked through a system of coordinate transformations. Surprisingly, this has seems to fail for Quantum mechanism, where the observer and the system being observed are linked so that the results of any observation is partly determined  by  the observer. This is indeed not so surprising if one consider the observer and the system being observed as \textit{entangled} quantum systems. This was a proposition of Everett \cite{everett} to introduce quantum observers in Quantum mechanics studies of systems, shifting from a \textit{probabilistic} to an \textit{unitary} formal theory. This has directly lead Albert and Loewer \cite{manyminds}, in their \textit{many-minds} interpretation to distinguish between the time evolution of an observer's physical state, which is continuous and unitary, and the evolution of an observer's mental state, which is discontinuous and probabilistic. Following these approaches, \textit{Universal quantum relativity} theory shall be developed based on the \textit{quantum information} theory since the assertions of any such theory have to do with the relationships between quantum system and observers treated as a coding problem integrating concepts of non-locality, uncertainty and complementarity of observables. 

In this article, we study the state of an entire informational universe  using a no-collapse interpretations of quantum mechanics combined with the postulate of the "absence of preferred basis". This allow to generalize the classical \textit{relational} point of view and to apply it to relations between the mind of the observer and a  empty scene surrounding it. Doing so, we enter an unknown territory in which space and time does not exist to begin with and where observers are entangled into an unique observer. In the theory, it is considered that spacetime is an emergent abstract background that may, in some approximations, replace relational observations by observations  in relation with this background. Thus spacetime only consists of \textit{elements of language} constrained by quantum theory. Moreover, among others, we consider gravity as non-fundamental, most likely an entanglement entropic force. In our model, the standard model of particles physics arises from entanglement effects while looking at a \textit{local realistic} theory.  It corresponds to a broken phase of the fundamental theory where the exclusion of some class of relative quantum states remains a good approximation.  This class of quantum states corresponds to those verifying criteria for violation of local realism.  

In section \ref{structure}, we present the formal structure of the theory. Then in section \ref{math}, we outline, the non-trivial underlying mathematical theory by describing the Hilbert space topology (\ref{hilbert}), the algebra of non-local observable (\ref{non-local}) and the algebra of local observable (\ref{local}). Finally in section \ref{discuss}, we present an interpretation of quantum mechanics, in the light of \textit{Universal Quantum Relativity}, that reconciles no-collapse interpretations and Copenhagen interpretation. 

\section{Theory formal structure}
\label{structure}
This section describes the formal structure of  \textit{Universal Quantum Relativity} in a logically isolated way.  As recommended by Steven Weinberg, a  \textit{Theory of Everything} should rely on the quantum theory together with a symmetry principle and a configuration space. We state that the three axioms below are sufficient to deduce all observable phenomena: 

\begin{enumerate}
\item first axiom of QM : \textit{State of a quantum system is represented by a vector in its Hilbert space}
\item symmetry principle:  \textit{Absence of absolute existence}
\item configuration space:   \textit{Infinitely divisible Universe quantum system}
\end{enumerate}

\noindent The first axiom of quantum mechanics is uncontroversial and hard to be seriously generalized keeping positive probabilities that sum up to unity. It implies superposition principle and unitary equivalence of all states in the Hilbert space which is the basic symmetry of quantum theory. 

The symmetry principle in the quantum framework may be rephrased as the “absence of preferred basis” i.e. laws of Physics must be the same in any computational basis. Any arbitrary introduced by the computational basis must be exactly compensate by a gauge freedom corresponding to the Hilbert space's automorphism. 

Regarding the configuration space, following Everett's \cite{everett} and Bohm's \cite{bohm} leads, we consider the state of the entire universe as a unique quantum system without external observer.  For such an indivisible Universe questions about outcomes cannot be even posed. The only way forward, in the absence of external reference point, is to use \textit{reference points within}, creations of the mind of the observer. Let's assume that the Hilbert spaces corresponding to this type of subsystems have two dimensions. This assumption is related to the ontological dichotomy \textit{"Is", "Is not"} and supported by law of parsimony. Without these two states, these two complementary and antagonist trends, these two ideas about things, no other ideas would exist. The Universe pure states in the simplest partitioned Hilbert space $\mathcal{H} = \mathcal{H}_{p_1} \otimes \mathcal{H}_{\mathcal{E}_{p_1}}$ reads

\begin{equation}
\ket{\psi} =  z_0 \ket{0}_{p_1}\ket{\psi_0}_{\mathcal{E}_{p_1}} + z_1 \ket{1}_{p_1}\ket{\psi_1}_{\mathcal{E}_{p_1}}
\label{yinyangeq}
\end{equation}

\noindent where we have used the Schmidt decomposition which is a consequence of a theorem of linear algebra that predates quantum theory\footnote{Without loss of generality we consider that states in $\mathcal{H}_{p_1}$ and in $\mathcal{H}_{\mathcal{E}_{p_1}}$ are defined only up to a global phase. Their relative phases with respect to the two terms of the sum has been then included in the generalized coefficient $z_0$, $z_1$ which become complex.}. It  demonstrates that any pure entangled bipartite state is a superposition of perfectly correlated outcomes of judiciously chosen measurements on each subsystem: Detecting $\ket{0}$ (resp. $\ket{1}$) on $p_1$ implies, with certainty, $\ket{\psi_0}$ (resp. $\ket{\psi_1}$) for $\mathcal{E}_{p_1}$, and \textit{vice versa}. This formulation of the Universe quantum state as entangled, in terms of relative states, is best represented  by the well known Taijitu diagram:

\begin{figure}[ht]
\centering
\includegraphics[scale=0.3]{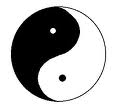}
\label{yinyang}
\end{figure}

\noindent This division process initiated in (\ref{yinyangeq}) can be pursue indefinitely. Namely, another \textit{reference points} $p_2$ may be considered in $\mathcal{E}_{p_1}$ leading to an environment $\mathcal{E}_{p_1 p_2}$. Ultimately, this iteration leads to a \textit{foam} of binary \textit{reference points within} $\mathcal{F}^\infty_O\equiv \mathcal{F}$ and its environment $\mathcal{E}_{\mathcal{F}^\infty_\mathcal{O}}\equiv \mathcal{E}_{\mathcal{F}}$  creation of the mind of the observer $O$. Following Universal Quantum Relativity principle, ambiguity in choosing computational basis of the composite Hilbert space $\mathcal{H}_{\mathcal{F}} \otimes \mathcal{H}_{\mathcal{E}_{\mathcal{F}}}$ has to be exactly compensate by a gauge freedom corresponding to its automorphism group.  

\section{Non-trivial underlying mathematical theory}
\label{math}

The information paradigm \textit{it from bit} introduced by John Archibald Wheeler has opened the door to theories where quantum theory sits alone at the foundation of Physics and that everything else, including spacetime and gravity,  is emerging from a quantum-information processing. \textit{Universal Quantum Relativity} implements this doctrine. We will show that quantum-information processing is possible using representation in anyone of the Cayley-Dickson algebras. Subsequent analysis makes use of singular, non-generic mathematical structure which is  the unique non-associative division algebra called octonions and the subtle $Spin(8)$ outer automorphism called \textit{triality}. 

\subsection{Hilbert space topology}
\label{hilbert}

\noindent As unitary equivalent quantum states are the fundamental entities of Universal Quantum Relativity, the topology of the partitioned Universe Hilbert space is of particular interest. According to the first axiom of quantum mechanics, the state of the entire Universe is represented by a vector $\ket{\psi}$ in its Hilbert space $\mathcal{H}$ defined up to a phase. This identifies the topology of $\mathcal{H}$ to the 1-dimensional sphere $S^1$: "One Spinning" or in the Latin word \textit{-Uni-}  and \textit{-Versus-}.  In order that from the "non-thing" arises the Whole, Universe quantum system can be partitioned  into subsystems through the \textit{division process} introduced in (\ref{yinyangeq}). For any integer $k\geq 1$, corresponding to a number of \textit{reference points within}, such as $\mathcal{F}^k_\mathcal{O}=\{ p_1,\cdots ,p_{k} \}$, the Universe quantum pure state can be represented as the sum of correlated outcomes labeled by bits sequences $s$ of length $k$:

\begin{equation}
\ket{\psi} = \sum_{s\in \mathbb F^k_2} z_s \ket{s}_{p_1\cdots p_{k}}\ket{\psi_s}_{\mathcal{E}_{p_1\cdots p_k}}
\label{foamk}
\end{equation}

\noindent The states labeled by bit sequences literally live in the mind of the quantum observer and  represent perfectly distinguishable states ($\braket{s}{s^\prime}=\delta_{ss^\prime}$) whereas correlated states, generally non-distinguishable, live in the observer environment.  \textit{The observer environment is an empty scene which does not have absolute existence nor properties such as dimensions.} The coherence factor  $\braket{\psi_s}{\psi_{s^\prime}}$ are arbitrary\footnote{This may be seen with a little thought from unitary condition in (\ref{foamk}) and $\braket{s}{s^\prime}=\delta_{ss^\prime}$.} and is characteristic from interference phenomenon. \textit{All of the physics comes from these interferences alone.} 

Unitary equivalence of all states implies that the topology of the partitioned Hilbert space $\mathcal{H}_{p_1} \otimes \cdots \otimes \mathcal{H}_{p_k}\otimes  \mathcal{H}_{\mathcal{E}_{p_1 \cdots p_k}}$ has to be identify with $S^{2^{k+1}-1}$ hypersphere as the sum over $s$ contains $2^{k}$ terms with complex coefficients. Remarkably, two consecutive ranks of representations following a division step are related by what we call the Taijitu equation:

\begin{equation}
\ket{\psi_s}_{\mathcal{E}_{p_1\cdots p_{k}}} = z_{s\vert 0} \ket{0}_{p_{k+1}}\ket{\psi_{s\vert 0}}_{\mathcal{E}_{p_1\cdots p_{k+1}}} 
 +z_{s\vert 1}\ket{1}_{p_{k+1}}\ket{\psi_{s\vert 1}}_{\mathcal{E}_{p_1\cdots p_{k+1}}}
\end{equation}
where $s\vert s^\prime$ represents two bits sequences concatenation.  

From a topological stand point, which is the basis of the study of Hilbert space automorphism, our division process appears closely related to the inductive doubling process that leads to the infinite family of finite-dimensional Cayley-Dickson algebras $A_k$. One can trade the usual Hypermatrix description of quantum propositions in $[\mathbb C^2]^k$ with matrix description with values in $A_k$. Also called hypercomplex algebras \cite{dickson}, the $A_k$, $k \in \mathbb{N}$ have dimension $2^k$ on $\mathbb{R}$. Each $A_k$ is the complexified of the previous algebras $A_{k-1}$, thanks to the existence of a vector in $A_{k-1}^{\perp}$, taken as the complex unity $\tilde{1}=(0,1)$. Hence, the topology of the partitioned Hilbert space $\mathcal{H}_{p_1} \otimes \cdots \otimes \mathcal{H}_{p_k}\otimes  \mathcal{H}_{\mathcal{E}_{p_1 \cdots p_k}}$ may be identify to $S^1_{A_k} \equiv \left \{(\alpha,\beta) \in A_k^2 \vert \alpha^2+\beta^2=1\right \}$. The consecutive partitioned Hilbert spaces present a nested \textit{bundle} structure that is mapped to an \textit{octonionic} structure.

\begin{equation}
\begin{array}{ccccccc}
\mathcal{H}_{\mathcal{F}^{k-3}} \otimes    \mathcal{H}_{\mathcal{E}_{\mathcal{F}^{k-3}}} & & \mathcal{H}_{\mathcal{F}^{k-2}}\otimes  \mathcal{H}_{\mathcal{E}_{\mathcal{F}^{k-2}}}& & \mathcal{H}_{\mathcal{F}^{k-1}}\otimes  \mathcal{H}_{\mathcal{E}_{\mathcal{F}^{k-1}}}  & & \mathcal{H}_{\mathcal{F}^{k}}\otimes  \mathcal{H}_{\mathcal{E}_{\mathcal{F}^k}}\\
&&&\\
S^1_{A_{k-3}} & \hookrightarrow & S^1_{A_{k-2}} & \hookrightarrow &  S^1_{A_{k-1}}&\hookrightarrow &  S^1_{A_{k}}\\
 && \downarrow && \downarrow & &\downarrow \\
 && A_{k-2}P^1 && A_{k-1}P^1 &&A_{k}P^1\\
\end{array}
\label{octonionuni}
\end{equation}

\noindent This bundle structure is well-defined up to $k=3$ as it concerns only the first four Cayley-Dickson algebras, i.e. $A_0=\mathbb{R}$, $A_1=\mathbb{C}$, $A_2=\mathbb{H}$, $A_3=\mathbb{O}$, which are viewed as well-behaved as being \textit {nicely normed, alternative, division}  algebras. In this case, one recovers the standard Hopf fibrations which is know in this context to be entanglement sensitive \cite{3qubits}. The base represents the state of \textit{one reference point within} together with the \textit{entanglement} with others. The larger Cayley-Dickson algebras are not alternative, contain non-trivial zero divisors\footnote{Indeed the limit of the Cayley-Dickson doubling process when $k\rightarrow\infty$ contains zero-divisors whose annihilator is ‘almost’ the whole algebra.}, and multiplication is anisometric.  This is usually a show stopper as it leads to \textit{logical paradox} \cite{Ak}\cite{QC}. However, as warned by Leibniz, privileging the logic corresponds to closing the door to inventiveness. The proposal to move forward is to use the theory of  "varying complexity representations of Cayley-Dickson algebras". The complexity of the structure of Cayley-Dickson algebra may be described using two sub-algebras.  Any $\alpha \in A_k$ may be written as

\begin{equation}\begin{split}
\alpha&=\sum_{i=0}^{2^{k}-1} \xi_i e_i \\
&= \sum_{j=0}^{2^{k-m}-1}\left(\sum_{l=0}^{2^{m}-1}\xi_{2^{m}j+l} e_l\right) \times e_{2^{m}j}\\
&=\sum_{j=0}^{2^{k-m}-1}\beta_j(\alpha) e_{2^{m}j}
\label{rep}
\end{split}\end{equation}
\noindent where $e_i$, $0\leq i \leq 2^{k}-1$  is the canonical base of $A_{k}$ and $\beta_j(\alpha) \in A_{m}$. The decomposition is unique. This structure depends solely on the multiplication table in $A_k$ where $e_{2^{m}j+l} = e_l  \times e_{2^{m}j}$ and is due to Eakin-Sathaye \cite{Eakin}. Some representations privilege \textit{logic} and emergence of \textit{meaning} whereas some representations privilege \textit{inventiveness of the calculation} and \textit{logical paradox} \cite{Ak}. 

For our purpose which is the study of the topology of the Hilbert space, let's  remark that for $m=k-3$ in (\ref{rep}), the $A_k$ displays an octonionic structure that link four level of algebras, if one identifies $A_{k-2}$ to $\mathbb{C}_{g_{k-2}}$\footnote{The kind of map allowing this identification will be made explicit for octonions in section \ref{sm}}. Then, a self-similarity based on recursion appears at the level of the Hilbert spaces \textit{bundle} structure in (\ref{octonionuni}). 

\begin{equation}
\begin{array}{cccccccc}
S_{\mathbb{R}}^1 & \hookrightarrow & S_{\mathbb{C}_{i_{k-2}}}^1 & \hookrightarrow & S_{\mathbb{H}}^1&\hookrightarrow &  S_{\mathbb{O}}^1 \\
& & \downarrow & & \downarrow& & \downarrow \\
& & \mathbb{C}_{g_{k-2}}P^1 & & \mathbb{H}P^1 & & \mathbb{O}P^1 \\
\end{array}
\label{octonionunideux}
\end{equation}

\noindent We call this feature \textit{Universal fractal}. We will see that this bundle structure has the richness to explain the standard model of elementary particle physics together with emerging spacetime and gravity. 

\subsection{About algebra of non-local observables}
\label{non-local}

We have seen in the previous section that states can be represented by matrix with values in $A_k$. The \textit{fundamental representation} of $A_k$ with \textit{maximal complexity} \cite{Ak} is obtained using coefficients in the largest alternative division algebra, the octonions, by setting $m=3$ in (\ref{rep}). In other words we have to use octonions to represent components if one wants to be able to \textit{divide}. This fact combined with the structure in (\ref{octonionunideux}) suggest that  \textit{Fundamental observables} of the composite Hilbert space $\mathcal{H}_{\mathcal{F}} \otimes \mathcal{H}_{\mathcal{E}_{\mathcal{F}}}$ can be faithfully represented by elements of the exceptional Jordan algebra $J_3^{\mathbb{O}}$ that is by Hermitian  $3 \times 3$ matrices with octonionic entries and trace 1 corresponding to points in the octonionic projective plane. Indeed the octonionic projective plane $\mathbb{O}P^2$ or \textit{Cayley-Moufang plane} is the only choice where one can explicit simultaneously the three possible hopf fibrations to $\mathbb{O}P^1$ as lines in $\mathbb{O}P^2$. Moreover $J_3^{\mathbb{O}}$ elements satisfy necessary algebraic conditions to represent projection operators associated to quantum mechanical states. Notably, $J_3^{\mathbb{O}}$ is compatible with Loop quantum gravity as it can reproduce Chern-Simons-like phases \cite{Jordan}. In addition $J_3^{\mathbb{O}}$ structure is known to code the degrees of freedom of string theory in a purely algebraic framework, under the form of  the 8 dimensional vector, spinor and conjugate spinors \cite{Jordan}.

In the absence of a preferred basis, only invariant quantities have a physical meaning. There is  a unique symmetric trilinear form $(.,.,.): J_3^{\mathbb{O}} \times J_3^{\mathbb{O}} \times J_3^{\mathbb{O}} \rightarrow \mathbb{R}$ such as $(\mathcal{A},\mathcal{A},\mathcal{A}) = det(\mathcal{A})$. $E_6$ transformations, interpreted as $SL(3,\mathbb{O})$,  leave invariant the determinant of $J_3^{\mathbb{O}}$  matrices. Among subgroup, $SO(8)$ or its double cover $Spin(8)$ plays a singular role as it preserves the diagonal of all $J_3^{\mathbb{O}}$ matrices \cite{E6}.  Indeed $Spin(8)$ group has an outer automorphism of order three that have the effect of permuting the vector, left-handed spinor, and right-handed spinor representations. This special features is called \textit{triality}. This symmetry is non-local in essence as it is relative to three distinct references. 

\subsection{About algebra of local observables}
\label{local}

\noindent If one wants to apply this formalism to particle physics, one must break the full $E_6=SL(3,\mathbb{O})$ symmetry. In our approach the breaking to $SL(2,\mathbb{O})$, $SL(2,\mathbb{H})$,  $SL(2,\mathbb{C})$ subgroups arise naturally while looking at local observables thanks to Hopf fibration as each base space in (\ref{octonionunideux}) is  related to a single reference. There is only three local topological patterns: $\mathbb{C}P^1$, $\mathbb{H}P^1$ and $\mathbb{O}P^1$. The automorphism groups of these patterns are precisely $SL(2,\mathbb{C})$, $SL(2,\mathbb{H})$ and $SL(2,\mathbb{O})$. There is a second level of breaking from $SL(2,\mathbb{O})$ and $SL(2,\mathbb{H})$ to $SL(2,\mathbb{C})$ that consists in isolating the entanglement contributions to local references.

\subsubsection{Triality breaking}

First, we ask about manifestation of \textit{triality} symmetry within the algebra of local observables. The three possible fibrations\footnote{Although the tensor-product nature of the $k$-references Hilbert space constrains the definition of $A_k^2$ from $[\mathbb{C}^2]^k$, there is still $k$ possibilities with the net effect to interchange the role of the $k$ references. This leads to $k$ different \textit{orientations} for the fibration whose base is associated to a single reference. } to the octonionic projective line  in (\ref{octonionunideux}) can be related to the fact that a matrix in $J_3^{\mathbb{O}}$ contains three local projections matrix of $J_2^{\mathbb{O}}$, corresponding to the three ways of embedding a $2\times2$  matrix inside a $3\times3$ matrix. Thus there is three distinct overlapping copies of $SL(2,\mathbb{O})$  in $SL(3,\mathbb{O})$. However,  there is only one  $SO(8)$ transformation, subgroup of the 3 copies of $SL(2,\mathbb{O})$, in an $E_6$ transformation \cite{E6}. Hence one can say that the fibrations down to local structure break the \textit{triality} symmetry. Reminiscence of \textit{triality} in the \textit{local world} are the  spinors and vectors interaction which is just the action of $J_2^{\mathbb{O}}$ on $\mathbb{O}^2$ by matrix multiplication. We consider that this is the correct formulation of Supersymmetry breaking in a \textit{Theory of Everything}. 

\subsubsection{Spacetime and classical gravity emergence}

Then, we consider the possibility of spacetime emergence. $SL(2,\mathbb{C})$  group is embedded in $SL(2,\mathbb{H})$ and $SL(2,\mathbb{O})$ groups. This shows that  \textit{The absence of a preferred basis} for the quantum Universe  is to be related to usual Lorentz symmetry associated to a single reference\footnote{This local reference may be composite if one has identified $A_{k-2}$ to $\mathbb{C}_{g_{k-2}}$ in (\ref{octonionunideux}).}. The so-called \textit{local Lorentzian references} are  \textit{elements of language} of the observer. Thus they are good candidates to represent a  dynamical abstract background with 3+1-dimension that we can call spacetime. Obviously, in our model local references store information. In the mean time, they are associated to $\mathbb{C}P^1\cong S^2$ topology in the Hilbert space. Following Penrose hint \cite{penrose}, we map $S^2$ on null ray over our emerging spacetime. We then assume that information is defined on null rays. This is compatible with an holographic description \cite{holo1} \cite{holo2}\cite{bekenstein} that allow to recover gravity as an entropic force \cite{Verlinde}\cite{lees}. Therefore, our approach is compatible with a thermodynamic interpretation of the field equations of gravity obtained from an entropy maximization principle. Let's add that all local states map inside a standard ball, where the set of separable states forms the $S^2$ boundary. Therefore it is mapped to the time-like region inside the light-cone. This suggest  an entanglement origins of \textit{inertia} in our emergent spacetime. Thus it offers the possibility to consider the Lorentz frame associated to a reference as a \textit{rest frame}. It is crucial to emphasis that the possibility to have a \textit{rest frame} associated to a reference is equivalent to the existence of a \textit{clock}. This is of course a mandatory step to be able to measure distance between \textit{local Lorentzian references}. 

\subsubsection{Standard model of elementary particle physics}
\label{sm}

Finally we have a look at the entanglement contribution on \textit{local Lorentzian references}. We have seen that $SL(2,\mathbb{C})$, $SL(2,\mathbb{H})$ and $SL(2,\mathbb{O})$ are the only possible realized local symmetries. Our strategy is to factorized the Lorentz transformation in order to isolate subgroup associated to entanglement. The guess of using Lorentz group as a preferred subgroup is not new in a dimensional reduction context \cite{dimred}. The difference here is that extra degrees of freedom are not related to any 10-dimensional or  6-dimensional spacetime.

First let's clarify our complex units. Quaternions are generated by two generators $g_0=i, g_1=j$ and a third generator $g_2=l$ generates octonions. The canonical basis such as $g_n=e_{2^n}$ reads $e_0=1, e_1=i, e_2=j, e_3=ij, e_4=l, e_5=ijl, e_6=il, e_7=jl$. So that any quaternion may be written as $q=\sum_{n=0}^3 q_ne_n, q_n \in \mathbb{R}$ and  any octonion may be written as $o=\sum_{n=0}^7 o_ne_n, o_n \in \mathbb{R}$. The maps 
\begin{equation}
\begin{array}{rcccl}
h_j(q)&=&1/2\left(q+jqj^*\right)&=&q_0+q_2j \\
h_l(o)&=&1/2\left(o+lol^*\right)&=&o_0+o_4l
\label{complexSub}
\end{array}
\end{equation}
project $\mathbb{H}$ and $\mathbb{O}$ to a complex subalgebra. Hence it breaks $SL(2,\mathbb{H})$ and $SL(2,\mathbb{O})$ symmetries to an  $SL(2,\mathbb{C})$ subgroup. We may also define the maps 

\begin{equation}
\begin{array}{rcccl}
t_j(q)&=&1/2\left(q-jqj^*\right)&=&q_1i+q_3ji \\
t_l(o)&=&1/2\left(o-lol^*\right)&=&o_1i+o_2j+o_3ij+o_5ijl+o_6il+o_7jl
\end{array}
\end{equation}
so that any quaternion or octonion have the decomposition $h+ t$. The first term  is called the \textit{head}. The second term is called the \textit{tail} and are \textit{doubly pure} quaternion or  octonion.  

We are now looking at the subgrougs $\mathcal{G}_1$ (resp. $\mathcal{G}_2$) such as $SL(2,\mathbb{C})\times \mathcal{G}_n$ transformations of $SL(2,\mathbb{H})$ (resp. $SL(2,\mathbb{O})$) preserve the $h+ t$ decomposition. $\mathcal{G}_1$ (resp. $\mathcal{G}_2$)  corresponds to transformations that preserve the multiplication table of the  \textit{doubly pure} quaternions (resp.  octonions).  In the case of the quaternions, such transformation are know to be its 3-dimensional automorphism group $SO(3)$. For the octonions, the transformation that preserve the multiplication table with one octonionic fixed units is known to be $SU(3)$ subroup of the exceptional group $G_2$. Thus $\mathcal{G}_1=SO(3)$ and $\mathcal{G}_2=SU(3)$.  Another way of envisioning this is to remark that $SL(2,\mathbb{C})\times \mathcal{G}_1$ (resp. $SL(2,\mathbb{C})\times \mathcal{G}_2$) are $\mathbb{H}P^1$ (resp $\mathbb{O}P^1$) automorphisms that have the 2-sphere $\mathbb{C}P^1$ as its fixed point set. $SL(2,\mathbb{C})\times \mathcal{G}_0$, where $\mathcal{G}_0=U(1)$,  has also  the 2-sphere $\mathbb{C}P^1$ as their fixed point set. Thus , up to some double-covering issues, factorizing the Lorentz transformation to isolate the symmetries associated with entanglement left us with $SU(3)\times SU(2)\times U(1)$ group.

To motivate this symmetry breaking, let's say that it is a mandatory step to consider entanglement effect locally if one wants to use a model with quantum field, as a collection of (infinitely many) quantum systems, each at every \textit{local Lorentzian references}.  Once the entanglement effect has been considered as internal, spinors and vectors interaction are represented by the action of $J_2^{\mathbb{C}}$ on $\mathbb{C}^2$ by matrix multiplication at every local Lorentzian references. Thus this approach  gives an explanation of the general features of the standard model of elementary particle physics. It explains why forces, except gravity, are described by a gauge theory with group $SU(3)\times SU(2)\times U(1)$. It provides a physical understanding on the origin of the split between Lorentz and internal symmetry. Internal symmetry origins in non-locality and appears as internal while looking at a local effective  theory. 

\section{Discussions}
\label{discuss}

In summary, we have adopted the point of view that reverses the usual classical notion that the independent elementary parts of the world and particular arrangements of these parts are the fundamental reality and rather say that inseparable quantum interconnectedness of the whole Universe is the fundamental reality \cite{bohm}. Moreover it has been understood that all the concepts we use to describe nature are not features of reality, but creations of the mind; parts of the map, not of the territory. The territory features all latent forms and power through the negative form which by not affirming anything allows all potentiality. This was summarized as an ultimate doctrine that is the \emph{absence of absolute existence}. Assuming  infinite divisibility of the Universe, this symmetry principle,  combined with the uncontroversial part of quantum mechanics, leads (without too much effort) to the known thermodynamic interpretation of the field equations of gravity and to the general features of the standard model of elementary particle physics. The symmetry breaking patterns used for the derivation have a clear explanation in terms of the observer's will to reach a \textit{local} comprehension of the fundamental non-local reality. 

A direction for future works is to exploit the additional non-local algebraic structure that we have highlighted as embedding the standard model structure. In particular, one shall look for the constraints from the non-local algebraic structure that lead to the local \textit{anomaly-free} theory. One shall check in details that the 'two parties' local topological pattern leads to a force that allow \textit{spontaneous symmetry breaking} and that the 'three parties' local topological pattern leads to a \textit{confining} force featuring \textit{asymptotic freedom}.  One shall also understand the relation between the \textit{Universal Fractal} feature and fermions \textit{flavors} and \textit{families}. Moreover, connections with algebraic description of the String theory \cite{Jordan} or black hole thermodynamics \cite{BH} shall be of particular interest. 

Last, but not least, we will also  need to better understand how the entirely quantum universe appears, for all practical purposes, to behave classically at the macroscopic level. In order to achieve this, we propose the following interpretation. We have seen that the distinguishable states labeled by bit sequences live in the separable mind of the unique observer represented by system $\mathcal{F}$. Correlated states or waves, generally non-distinguishable, live in the observer environment $\mathcal{E}_\mathcal{F}$ relative to $\mathcal{F}$. Usually, in no-collapse interpretations of quantum mechanics, the various \textit{many} correlated outcomes, that sum-up to the Universe wave function, would lead to either \textit{many-world} or \textit{many-mind} interpretation. We'd rather propose again an analogy with Holography.  The superposition of two or more waves in $\mathcal{E}_\mathcal{F}$ interfere to produce new waves (or wavefronts). \textit{All that can be observed}, are these interferences. This may be thought as \textit{fundamental spontaneous collapses} generating the hologram or world we have the experience of. The following picture describe in details the proposed analogy with holography.
\begin{figure}[ht]
\centering
\includegraphics[scale=0.5]{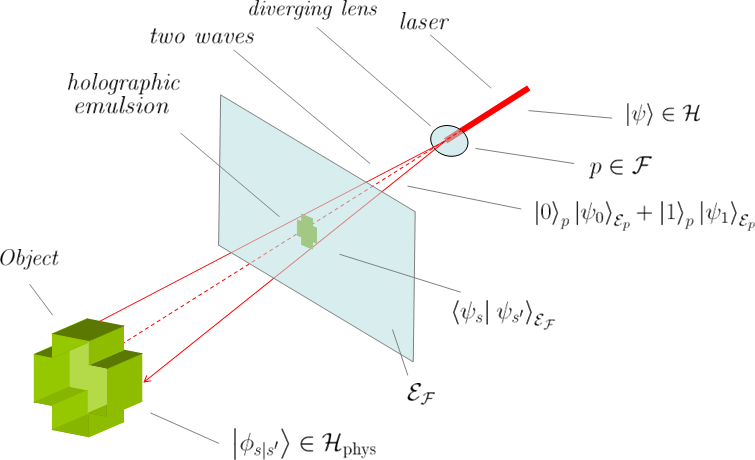}
\label{holo}
\end{figure}

 One observes wavefront recorded in $\mathcal{E}_\mathcal{F}$ that are collectively created. The observer needs at least a quantum binary \textit{reference point} $p$ in his mind, possibly composite, to see. Without these two ideas about things, no other ideas would exist. Then what is observed may be interpreted as quantum wavefront $\ket{\phi_{s|s^\prime}}$ associated to an \textit{open} quantum systems, living on Lorentzian local references. Note that the reference point $p$ is part of the system $\mathcal{F}$. Therefore the observation is partly determined by the observer in inverse proportion to the length of the bit sequences $s$ and $s^\prime$. Information redundancy give rise to \textit{objective existence}. 

As we have been doing global assumptions, implications of this work have also a cosmological meaning. Using the decoherence paradigm \cite{superselection}, we would expect a  decay or vanishing of the coherence factors $\braket{\psi_s}{\psi_{s^\prime}}$ which code the wavefronts of the hologram. This gives a direction to explain the problem of time in cosmology and the phenomena of an \textit{observed Universe} in accelerated expansion. 

\section*{Acknowledgement}

\noindent I wish to thank Christian Korthals-Altes for being my advisor during my PhD a little bit more than a decade ago, Michel Palosse for useful discussions during the redaction of this article, Master Zhu for encouraging me in "proving the yin-yang". I also greatly appreciate the support and encouragement of my wife, Bernadette and of my Children Eleonore, Benjamin  and Jean-Baptiste. A special thanks to Neale Donald Walsch's conversations that helped me to \textit{re-member} and to Frifjof Capra for his inspiring "out-of-the-box" approach to physics. Thanks to http://arxiv.org/  and http://en.wikipedia.org/ for the access to information. Om Namah Shivaya.

\end{document}